\documentstyle[12pt]{article}
\begin{document}

\begin{titlepage}

\begin{flushright}
\vbox{
{\bf TTP93-11}\\
{\bf March 1993}\\
 }
\end{flushright}

\vskip 1.8cm
\begin{center}
{\bf\large
MOMENTUM DISTRIBUTIONS IN ${\bf t\bar t}$}\\
{\bf\large PRODUCTION AND DECAY}\\
{\bf\large NEAR THRESHOLD (II):}
\end{center}
\begin{center}
{\bf\large MOMENTUM DEPENDENT WIDTH}\\
{\it (Extended version of preprint TTP92-38)}
\end{center}
\vskip 1.5cm
\begin{center}
{\bf\Large M.~Je\.zabek \footnote{Alexander von Humboldt
Foundation Fellow. Permanent address:
Institute of Nuclear Physics, Krak\'ow, Poland;
e--mail: jezabek@chopin.ifj.edu.pl and bf08@dkauni2.bitnet}
and T.~Teubner \footnote{e--mail:
teubner@physik.uni-karlsruhe.dbp.de}}
    \\ \vskip0.3in
   {\it Institut f\"ur Theoretische Teilchenphysik}\\
   {\it Universit\"at Karlsruhe}\\
   {\it Kaiserstr.~12, Postfach 6980}\\
   {\it 7500 Karlsruhe 1, Germany}
\end{center}
\vskip1.0cm

\begin{center}
{\bf Abstract}
\end{center}
\begin{quote}
{\small\rm  We apply the Green function formalism for
$t-\bar t$ production and decay near threshold in a study
of the effects due to the momentum dependent width for
such a system. We point out that these effects are likely
to be much smaller than expected from the reduction of the
available phase space. The Lippmann--Schwinger equation for
the QCD chromostatic potential is solved numerically
for $S$ partial wave. We compare the results on the total
cross section, top quark intrinsic momentum distributions
and on the energy spectra of $W$ bosons from top quark decays
with those obtained for the constant width.}
\end{quote}

\end{titlepage}
\mbox{}
\thispagestyle{empty}
\newpage


\section{Introduction}
\setcounter{page}{1}
There are accumulating indications that the long awaited top
quark will be discovered soon. The value of its mass $m_t$
preferred by the indirect LEP measurements \cite{JEllis}
implies that the $t$ quark is a short--lived particle, and
its width $\Gamma_t$ is of the order of several hundreds MeV.
As a consequence the cross section for $t\bar t$ pair
production near energy threshold has a rather simple and
smooth shape. In particular, it is likely that in $e^+e^-$
annihilation only the $1S$ peak survives as a remnant of
toponium resonances \cite{Workshop,TQPTA}\,.
Nevertheless, as pointed out by Fadin and Khoze
\cite{FK} the excitation curve
$\sigma(e^+e^-\rightarrow t\bar t\,)$
allows a precise determination of $m_t$ as well as of
other physical quantities such as $\Gamma_t$ and the strong
coupling constant $\alpha_s$\,. These results, derived
analytically for the Coulomb chromostatic potential, have
been confirmed by Strassler and Peskin \cite{SP}
who studied numerically a more realistic
QCD potential. The idea \cite{FK,SP} to use Green function
instead of summing over overlapping resonances has been
also applied in calculations of differential cross sections:
energy spectra of $W$'s from decays \cite{JKT}~,
and intrinsic momentum distributions of top quarks
in $t\bar t$ systems \cite{JKT,Sumino}\,.\\

Pandora's box was opened when the authors of \cite{Sumino}
found that the effects of the momentum dependent width may
show up in the annihilation cross section
$\sigma(e^+e^-\rightarrow t\bar t\,)$~. Despite the fact
that the calculation is plagued by unpleasant and
unphysical gauge dependence, one has to consider their
conclusions as an indication that the accuracy of the
calculations \cite{FK,SP,JKT} assuming a momentum
independent decay rate should be reconsidered. One cannot
use the results of \cite{Sumino} for such an
estimation, because some potentially relevant effects
are not taken into account there.\\

In the present paper we extend the calculation
of \cite{JKT} by including a momentum dependent width.
Our work is based on an observation \cite{Huff} that
relatively simple semiclassical arguments reproduce
well the results of a complete calculation
of the lifetimes and the energy spectra for decays of
muons bound in light and medium nuclei. Since the strength
of interaction is comparable for $t\bar t$ and
$\mu^-$--nucleus for the nucleus charge
$Z \sim$~10--20, one may expect that analogous
considerations closely approximate the results of the
complete QCD calculations for
top quark pair production near threshold.\\

One of the most important future applications of this
calculation is the determination of $m_t$ and $\alpha_s$
from the combined measurements of the total and the differential
cross sections in $e^+ e^- \to t\bar t$\,. We study the
theoretical uncertainties related to model assumptions on
the momentum dependent width. These result from
${\cal O}(\alpha_s^2)$ corrections to the process
$e^+ e^- \to t\bar t \to b W^+ \bar b W^-$ which have
been not calculated yet. Some corrections due to reduction
of the available phase space \cite{JK}
have been considered in \cite{Sumino}. However, as we
already pointed out in our previous paper \cite{JKT} one
should expect large cancellations between the corrections
due to the phase space reduction and the Coulomb enhancement
of the $b$ and $\bar b$ wave functions. In order to determine
theoretical uncertainties related to the momentum dependent
width we evaluate predictions of a model ({\it Model I}\,) which
overestimates the effects on $\sigma(e^+e^- \to t \bar t\,)$ and
the differential cross sections. Nevertheless, the resulting
theoretical uncertainties in determination of $m_t$ and
$\alpha_s$ are quite small.\\
We also work out the predictions of a more realistic model
({\it Model II}\,) for the momentum dependent width and find
out that for this model the results are close to those
obtained assuming constant width.\\

Our paper is organized as follows: In sect.2 we describe the
physics of bound $t\bar t$ decays and formulate our models
of the momentum dependent width. In sect.3\,, which we add for
the sake of completeness, the Green function method and our
QCD potential are briefly described. In sect.4 the
results of the numerical calculations are presented, and in
sect.5 our conclusions are given.

\section{Models for the width of ${\bf t-\bar t}$ system}

In our earlier paper \cite{JKT} we followed \cite{FK,SP} and
confined the discussion to the non--relativistic
approximation and the constant decay rate.
The basic reason was that numerous effects related to the
intrinsic momentum in the $t -\bar t$ system were known to
be small and they were contributing with opposite signs to
the width. Let us list now the main physical effects which
result in the momentum dependent width.
The width of the $t-\bar t$ system depends on
the intrinsic momentum, of say $t$ quark, because both the
matrix element and the phase space available for the decay
products depend on it. The phase space effect tends to
reduce the decay rate of bound top quarks relative to free
ones \cite{JK}, and the effect is enhanced, because for
short--lived particles the distribution of intrinsic
momentum is broad. However, for the same reason the decays
take place at short relative distances, where the wave functions
of $b$ and $\bar b$ quarks originating from the decays are
distorted (enhanced) by Coulomb attraction. Therefore, when
calculating the amplitude of $t\rightarrow bW$ transition,
one should use Coulomb wave functions rather than plane waves
for $b$ quarks. This effect clearly increases the rate. A third
factor is the time dilatation: a top quark moving with velocity
$v$ lives longer in the center--of--mass laboratory
frame\footnote{In an analogous calculation for $\mu^-$ bound in
a nucleus of charge $Z$ one obtains \cite{Huff}
$$\Gamma = \Gamma_{free}
\left[1-5(Z\alpha)^2\right]\left[1+5(Z\alpha)^2\right]\left[
1-(Z\alpha)^2/2\right]$$ where the first correction factor
comes from the phase space suppression, the second from the
Coulomb enhancement, and the third one from time dilatation.}.
Let us consider now QCD corrections to the decay
$t\rightarrow bW$~. The calculations of the corrections to
the total rate \cite{JK1}~, the energy spectrum of the charged
lepton \cite{JK2} and the energy of $W$ \cite{CJK1} do not
include two effects which may be non--negligible for
$t-\bar t$ production near threshold:
a) the QCD Chudakov effect, i.e.~the suppression of soft gluon
emission from a small and colour singlet $t\bar t$ source, and
b) the additional suppression of the transition from a colour
singlet into a colour octet $t\bar t$ system due to the related
change of its potential energy\footnote{For a recent discussion
of the effect of $\Gamma_t$ on soft gluon
radiation see \cite{ODKS}.}. In the following we ignore all these
effects, as well as analogous effects in transverse gluon QCD
correction to the production vertex. Since ${\cal O}(\alpha_s)$
QCD corrections to the decay \cite{JK1,JK2,CJK1,CJK2}
are $\sim$10\%, we
hope that the accuracy of our approximation is of the order of 1\%,
i.e.~that it is comparable to the size of ${\alpha_s}^2$ effects
which are also neglected. In particular we neglect all relativistic
and spin corrections but the time dilatation in the momentum
dependent width. QCD corrections to the decay rate
$\Gamma_{t-\bar t}$ are included as an overall
factor $f$ modifying the width $\Gamma_t$ for single top decay:
\begin{equation}
\Gamma(p,E)\; = \;{1\over 2}\Gamma_{t-\bar t}\; = \;{\cal C}(p,E)
\Gamma_t
\label{eq:II1}
\end{equation}
where
\begin{equation}
\Gamma_t =
\frac{G_Fm^3_t}{8\sqrt 2 \pi}
\left( 1 - \frac{m^2_W}{m^2_t}\right)^2
\left( 1 + 2 \frac{m^2_W}{m^2_t} \right)
\left[ 1 - \frac{2}{3} \frac{\alpha _s}{\pi} f({m^2_W}/
{m^2_t})\right]
\label{eq:II1a}
\end{equation}
and \cite{JK1}
\begin{eqnarray}
f(y) =
 \pi ^2 + 2 Li_2 (y) - 2Li_2 (1-y)
+ [ 4y(1-y-2y^2) \ln y + 2(1-y)^2{\bf \cdot}
\nonumber \\
(5+ 4y) \ln(1-y)
 - (1-y)(5+9y-6y^2)] /
[ 2(1-y)^2 (1+2y) ] 
\label{eq:II2}
\end{eqnarray}
\par\noindent
The correction factor ${\cal C}(p,E)$ depends on the intrinsic
momentum $\vec p$ and the energy $E=\sqrt{s}-2m_t$ as
explained at the beginning of this section. Some effects such as
the phase space reduction or the time dilatation can be easily
implemented. In fact these effects have been already included
in the calculations performed in \cite{Sumino}. The Coulomb
enhancement cannot be easily taken into account. In principle
one has to replace the plane wave functions for $b$ quarks by
relativistic Coulomb functions when evaluating the amplitude
for $t\rightarrow bW$ transition. One may hope, however, that
the following observation, valid for muons bound in nuclei
\cite{Huff}, holds also for chromostatic attraction in
$t-\bar t$ systems: the phase space suppression and the Coulomb
enhancement nearly cancel each other. For light nuclei the result
is well described by the time dilatation suppression alone. For
heavier nuclei there is a further reduction in the decay rate
because some of the decay electrons will not have sufficient
kinetic energy to escape to infinity from the region of the
electrostatic potential of the nucleus. We think that for
$t-\bar t$ it is reasonable to replace this `escape into infinity'
condition by a condition on the effective mass $\mu_{b\bar b}$
of the $b-\bar b$ system resulting from the decays of $t$ and
$\bar t$.\\
At first we discuss a Model I which overestimates the effects of
the momentum dependent width. Comparing the suppression factors
corresponding to Model I with the ones discussed in \cite{Sumino},
one observes that they are not very different, see
Fig.~\ref{fig.1}b in the present paper and
Fig.~6 in \cite{Sumino}. Thus, the results derived from
Model I should resemble the ones obtained in \cite{Sumino}.
We assume that for a $t-\bar t$ system
the $t$ quark with the intrinsic three--momentum $\vec p$ decays
in its rest frame as a free particle of mass $m_t$\,. $W^+$
originates from this decay with energy fixed by two--body
kinematics. Then its four--momentum in the $t-\bar t$ rest frame
is obtained from the Lorentz transformation. The same procedure
is applied to $W^-$ from the decay of $\bar t$. $W$'s from the
decays are colour singlets, so they escape without any final
state interaction. On the other hand the $b$ and $\bar b$
interact and their momenta are changed by final state
interactions. The four--momentum of the $b-\bar b$
system is fixed by energy--momentum conservation. It is clear
that for such a model the energy spectra of $W$'s are
broadened due to the Fermi motion. The effective mass
$\mu_{b\bar b}$ can be smaller than for decays of unbound $t$
and $\bar t$. For some configurations of the momenta of $W$'s
the resulting $\mu_{b\bar b}^2$ can even be negative. These
configurations must be rejected. We impose a stronger condition
on $\mu_{b\bar b}$ and require that $\mu_{b\bar b}$
must be larger than $\mu_0 = 2m_b+\Delta$~,
where $\Delta$= 2 GeV.  It is plausible that if this condition
is fulfilled the $b-\bar b$ system  decays into hadrons with
probability one. For $\mu_{b\bar b}<\mu_0$ we assume complete
suppression. Thus, our Model I for the momentum dependent width
is defined by the following formula:
\begin{equation}
{\rm d}\Gamma(\vec p,E) = \Gamma_t\,
\sqrt{1-{\vec p\,^2/( \vec p\,^2+{m_t}^2)}}\;
\Theta\left(\mu_{b\bar b} -\mu_0\right)\,{\rm d}L_{2\otimes2}
\label{eq:II3}
\end{equation}
where
the volume of the phase space $L_{2\otimes2}$
for $(bW^+)(\bar bW^-)$ is normalized to one. Thus, if the
suppression factors were absent the width $\Gamma_{t-\bar t}$ of
the $t-\bar t$ system would be equal to the sum of the widths
of free $t$ and $\bar t$. The first suppression factor
in (\ref{eq:II3}) is due to the time dilatation, and
the step $\Theta$ function describes the additional
suppression.
\begin{figure}
\vskip 1in
\vskip -0.2cm
\caption{{\it The suppression factors ${\cal C}(p,E)$ for
the decays of $t-\bar t$ system
in comparison to the single quark decays
of $t$ and $\bar t$: a) $m_t = 120$ GeV and b) $m_t = 150$ GeV.
Model I:
the energies correspond to $1S$ peak --- dashed, $E=0$ --- dotted,
and $E=2$ GeV --- dashed--dotted lines. Model II: energies of $1S$
peak ---
solid lines.}}
\label{fig.1}
\end{figure}
Depending on the total energy and the intrinsic momentum of the
decaying top quarks, the above--mentioned effects
result in a decrease of the width
$\Gamma_{t-\bar t}$.
In Figs.~\ref{fig.1}a--b
the correction factors ${\cal C}(p,E)$
are plotted for the top
masses $120$ and $150$\ GeV, respectively, and for three energies
near the threshold.
The phase space reduction
is larger for energies below threshold and leads together with the
relativistic time dilatation factor to a
strong suppression of decays for large values of the intrinsic
momentum. Both effects also depend on the top mass: the
larger $m_t$~, the smaller the reduction of $\Gamma_{t-\bar t}$.
The energy dependence is also weaker for smaller values of
$\Delta$~. However, the latter effect is rather small.\\
Let us describe now Model II which we consider to be more realistic.
Instead of (\ref{eq:II3}) the momentum dependent width is given by
the formula:\pagebreak[4]
\begin{equation}
{\rm d}\Gamma(\vec p,E) = \Gamma_t\,
\sqrt{1-{\vec p\,^2/( \vec p\,^2+\mu^2)}}\;
\Theta\left(\mu_{b\bar b} -\mu_0\right)\,{\rm d}L_{2\otimes2}
\label{eq:II4}
\end{equation}
where
\begin{displaymath}
\mu^2 = p_{\alpha} p^{\alpha}
\end{displaymath}
and $p_{\alpha}$ is the four--momentum of the off--shell top quark.
In its rest frame the energy of $W$ originating from the decay is
given by the two--body kinematics:
\begin{equation}
E_W^{\ast} = \frac{\mu^2 + m_W^2 - m_b^2}{2 \mu}
\label{eq:II5}
\end{equation}
The correction factor ${\cal C}(p,E)$ evaluated from Model II
depends weakly on energy and is close to $1$, see solid lines in
Figs.~\ref{fig.1}a--b.
Therefore , the corresponding results are much closer to those
derived assuming constant width.
Two more remarks are in order here.\\
The models
which we consider are inspired by the observation \cite{SZ}
that for short--lived top quarks the final states hadrons
originate from two--jet $b-\bar b$ configurations. The lower the
value of top mass and the longer its lifetime, the less realistic
become our assumptions. For small $\Gamma_t$ the decays take
place in a well--defined sequence, and the treatment of the
recoil effects given in \cite{JK} is more appropriate.\\

The width for off--shell top quarks is a gauge dependent object.
The results for physical quantities do not depend on the choice
of gauge because the gauge dependent contributions from
propagators, vertex corrections, wave function renormalization
and non--resonant graphs cancel order by order in perturbation
theory. The problem of gauge dependence in $Z^0$ and $W$ pair
production has been recently discussed in the literature
\cite{stuart}. The case  of $t\bar t$ production near threshold
is even more complicated because of multiple gluon exchange
and resonance formation due to Coulomb--like strong attraction.
In our opinion the only way to avoid gauge dependence in final
results is to solve the bound state problem including the
constant on--shell width, and to treat the difference between
on--shell and off--shell widths in the
framework of perturbation theory. However, such a complete
approach is beyond the scope of the present paper. Our main goal
is to estimate the importance of the effects related to the
momentum dependent width. Therefore, we follow \cite{Sumino} and
include the momentum dependent width but ignore all other
contributions. We argue that due to the
cancellations between phase space suppression and Coulomb
enhancement the difference between the on--shell and off--shell
widths is strongly reduced. Thus, the related differences in the
cross sections are small, and their gauge dependence may be
unimportant from the practical point of view.

\section{Green function for ${\bf t-\bar t}$ system}

In this section we briefly describe the Green function method
for $e^+e^-\rightarrow t\bar t$ annihilation and
the numerical solution of the Lippmann--Schwinger equation
near the energy threshold; see \cite{JKT} for details.
In our discussion we neglect $Z^0$ contribution and transverse
gluon correction to the production vertex.
The differential cross section
for top quark pair production in electron positron annihilation
reads \footnote{In (7) of \cite{JKT} a factor $(2\pi)^{-3}$
is omitted. The results, plots and conclusions presented there
are not affected by this omission.}
\begin{equation}
{{\rm d}\sigma\over {\rm d}^3p} \left(\vec p,E\right) =
{3\alpha^2\,Q_t^2\over\pi\,s\, m_t^2}
\Gamma(p,E)\left|{\cal G}(\vec{p},E)\right|^2\ .
\label{eq:III1}
\end{equation}
The Green function ${\cal G}(\vec{p},E)$ is the solution of
the Lippmann-Schwinger equation
\begin{equation}
{\cal G}(\vec{p},E) =
{\cal G}_0(\vec{p},E) +
{\cal G}_0(\vec{p},E)
\int {{\rm d}^3q\over(2\pi)^3}
\tilde V(\vec{p}-\vec{q}\,)
{\cal G}(\vec{q},E)
\label{eq:III2}
\end{equation}
where $\tilde V(\vec{p}\,)$ is the potential in momentum space.
The free Hamiltonian that is used to define the Green function
${\cal G}_0$
includes the momentum dependent width:
\begin{equation}
{\cal G}_0(\vec{p},E)
= {1\over E- {p^2\over m_t}+ {\rm i}\Gamma(p,E)}
\label{eq:III8}
\end{equation}
In our numerical calculations we use the two--loop QCD potential
\cite{potential}
for $n_f=5$ quark flavours, and join it smoothly to a
Richardson--like phenomenological potential \cite{Richardson}
for intermediate and
small momenta. This potential gives a very good description
of $J/\psi$ and $\Upsilon$ families.\\
We define the potential $\tilde V(p)$, where $p=|\vec p\,|$
as follows:
\begin{equation}
\tilde V_{JKT}(p) = - \frac{16\pi}{3}
\frac{\alpha_{eff}(p)}{p^2} +
V_0\,\delta(p)
\label{eq:III3}
\end{equation}
where
\begin{eqnarray}
\alpha_{eff}(p)=\left\{
      \begin{array}{ll}
       \alpha_{pert}(p) & \qquad {\rm if}\ p>p_1\ ,\\
       \alpha_R(p)+(p-p_2)\frac{\alpha_{pert}(p_1)-
        \alpha_{pert}(p_2)}{p_1-p_2} & \qquad {\rm if}\
       p_1>p>p_2\ ,\\
       \alpha_R(p) & \qquad {\rm if}\ p<p_2\ .\\
      \end{array} \right.
\label{eq:III4}
\end{eqnarray}
with
\begin{equation}
\alpha_{pert}(p) = \alpha_{\overline{MS}} (p^2,n_f=5)
  \left[ 1 + \left( \frac{31}{3} -
  \frac{10}{9} n_f \right)
   \frac{\alpha_{\overline{MS}}(p^2)}{4\pi} \right]
\label{eq:III5}
\end{equation}
and
\begin{eqnarray}
\alpha_{\overline{MS}}(p^2) &=&
\frac{4\pi }{b_0 \ln \left( p^2/
{\Lambda^{(n_f)}_{\overline{MS}}}\,^2
\right)
 + \frac{b_1}{b_0} \ln\ln
          \left(  p^2 /
{\Lambda^{(n_f)}_{\overline{MS}}}\,^2\right) }
\label{eq:III6}\\
b_0 &=& 11 - \frac{2}{3}n_f\,, \qquad
b_1 = 102 - \frac{38}{3}n_f\,.
\nonumber
\end{eqnarray}
The Richardson--like part is given by
\begin{equation}
\alpha_R(p) = \frac{4\pi}{9} \left[ \frac{1}
{\ln(1+p^2/\Lambda_R^2)} -
 \frac{\Lambda_R^2 q_{cut}^2}{p^2(p^2+q_{cut}^2)} \right]\,.
\label{eq:III7}
\end{equation}
Throughout our calculations we take $p_1=5$ GeV, $p_2=2$ GeV,
$\Lambda_R=400$ MeV and $q_{cut}=50$ MeV\,.
The constant $V_0$ is fixed by the requirement that
after Fourier transform the potential in the position space
\begin{displaymath}
V_{JKT}(r=1 {\rm GeV}^{-1}) = - 1/4 {\rm GeV}\,.
\end{displaymath}
\begin{figure}
\vskip 1in
\vskip -0.2cm
\caption{{\it $\alpha_{eff}(q)$ for different values of
$\alpha_s(M_Z)$: solid: 0.12, dashed: 0.11, dashed--dotted:
0.13, dotted: 0.10 and 0.14\,.}}
\label{fig.2}
\end{figure}
\begin{figure}
\vskip 1in
\caption{{\it QCD--Potential in the position space $V_{JKT}(r)$
for different values of $\alpha_s(M_Z)$:
solid: 0.12, dashed: 0.11, dashed--dotted: 0.13, dotted:
0.10 and 0.14\,.}}
\label{fig.3}
\end{figure}
In Fig.~\ref{fig.2} we plot $\alpha_{eff}$ for $\alpha_s(M_Z)$
varying between $0.10$ and $0.14$ whereas in Fig.~\ref{fig.3}
our potential $V_{JKT}(r)$ in the position space is shown.\\
Near the energy threshold one can neglect all but $S$ partial
waves and numerically solve the corresponding one--dimensional
integral equation. The spherically symmetric
solution fulfills the unitarity condition
\cite{Sumino}
\begin{equation}
\int \frac{{\rm d}^3p}{(2\pi)^3}\,
\Gamma(p,E)\,\left|{\cal G}(p,E)\right|^2 =
- {\rm Im} G(\vec x=0,\vec x\,^{\prime}= 0,E)
\label{eq:III9}
\end{equation}
which for the constant decay rate reduces to
the formula for the total cross section  derived
in \cite{FK}.\par

\section{Results}

As desribed in the last section, the momentum dependent width
\begin{figure}
\vskip 1in
\vskip -0.2cm
\caption{{\it Comparison of the differential cross sections
${\rm d}\sigma\over {\rm d}p$ evaluated for the momentum
dependent $\Gamma(p,E)$
(Model I: dotted, Model II: solid)
and constant $\Gamma_t$ widths (dashed lines) for $m_t=120$~GeV;
and a)$E=-2.3$, b)$E=0$ and c)$E=2$~GeV, respectively.}}
\label{fig.4}
\end{figure}
$\Gamma(p,E)$ enters
our numerical calculation of the Green function
${\cal G}(p,E)$.
The differential cross sections
${{\rm d}\sigma\over {\rm d}p}$ derived from (\ref{eq:III1})
are different from the analogous cross sections obtained
assuming constant width.
In the latter case ${1\over\Gamma_t}{{\rm d}\sigma\over {\rm d}p}$
is proportional to
$|p{\cal G}(p,E)|^2$\,, and we show these distributions
as dashed lines in Figs.~\ref{fig.4}a--c
for $m_t$ = 120 GeV  and $\Gamma_t$ = 0.3 GeV~.
For a comparison we plot also analogously normalized
cross sections
${\Gamma(p,E)\over\Gamma_t}
|p{\cal G}(p,E)|^2$
for the momentum dependent width
(dotted lines for Model I and solid for Model II).
The main
effect of the varying width $\Gamma(p,E)$ in Model I is a change
in the normalization
of the distribution ($\approx 15\%$ for the $1S$ peak)
accompanied
by the expected suppression of the large momentum tail. Thus the
Green functions are slightly shifted to smaller momenta.
However, it is remarkable that the position of the maximum is not
much affected.
\begin{figure}
\vskip 1in
\vskip -0.2cm
\caption{{\it Comparison of the annihilation cross sections
$\sigma(e^+e^- \to t\bar t\,)$ (in units of R) evaluated for
the momentum dependent $\Gamma(p,E)$ (Model I: dotted,
Model II: solid) and constant $\Gamma_t$ widths (dashed lines).}}
\label{fig.5}
\end{figure}
The energy dependence of the annihilation
cross section in the threshold region
can be obtained by integration of the
differential cross section over the intrinsic momentum:
\begin{equation}
\sigma(e^+e^- \to t\bar t\,) |_E = \frac{16\alpha_{QED}^2}
{3sm_t^2}
\int_0^{\infty}{\rm d}p\ p^2 \Gamma(p,E) |{\cal G}(p,E)|^2
\label{eq.res.1}
\end{equation}
In Figs.~\ref{fig.5}a--b we show $\sigma(e^+e^- \to t\bar t\,)$
as functions of energy for two values of $m_t$~: $120$ and
$150$ GeV. Once again we observe a stronger effect of the
momentum dependent width for smaller $m_t$~:
while for a top mass of $120$~GeV an effective decrease of
the width in Model I leads to the increase of the cross section
at $1S$ resonance by more than $15\%$, the
total cross section is much less affected
for top masses as high as  $150$~GeV and larger.
Above the $1S$ peaks, and in particular above the threshold,
the effects of the momentum dependent width are small, typically
of the order of few percent.
We observe that also for the total cross section the position of
the $1S$ peak is not shifted, despite the fact that in Model I
the normalization is significantly changed. It can be shown
\cite{MMO} that the position of the peaks in the total
$\sigma(e^+e^- \to t\bar t\,)$
and differential $\frac{{\rm d}\sigma}{{\rm d}p}$ cross sections
are essential for a precise simultaneous determination of $m_t$
and $\alpha_s$ from the threshold region. Thus we conclude that
the precision of such an analysis is not affected by theoretical
uncertainties related to the momentum dependent width.\\

As expected, the results obtained from Model II are close (within
a few percent) to the results corresponding to the constant
width, c.f.~solid and dashed
lines in Fig.~\ref{fig.4} and \ref{fig.5}.\\

\begin{figure}
\vskip 1in
\vskip -0.2cm
\caption{{\it Normalized energy spectra of $W$ bosons
originating from the decays of $t-\bar t$ systems at $1S$
peaks (narrower) and for $E=2$ GeV (broader). The dashed lines
correspond to the momentum dependent width evaluated in Model
I and solid ones to the constant widths.
The $W$ spectra obtained from Model II are very close to the
solid curves (constant width case).}}
\label{fig.6}
\end{figure}
The top momentum distributions themselves cannot
be measured experimentally. They can be reconstructed
kinematically from the four--momenta of the final state
particles, but, as should be clear from our discussion,
those are sensitive to model assumptions. In particular the
four--momenta of $b$
and $\bar b$ quarks are distorted by final state interactions.
The final state interactions, however, do not change the
four--momenta  of $W$'s which are colourless. Thus their
distributions contain important information on the dynamics
of production and decay of top quarks. In Figs.~\ref{fig.6}a--b
we compare the normalized energy spectra of $W$'s
calculated for the momentum dependent
$\Gamma(p,E)$ in Model I and constant $\Gamma_t$ widths.
The calculations have been performed for
the non--zero  width of $W$ bosons,
$\Gamma_W$ = 2.2 GeV. Since the distributions of the intrinsic
momentum are narrower for $\Gamma(p,E)$, c.f.~Figs.~\ref{fig.4}
and \ref{fig.5}, the resulting energy spectra of $W$'s are also
narrower than the corresponding distributions for the constant
width. We show these distributions as dashed and solid
lines for the varying and constant width, respectively.
The sharper spectra are obtained
for the energies corresponding
to $1S$ peaks and the broader ones for $E=2$ GeV. The
spectra of $W$'s obtained from Model II are close to those for
the constant width.\\

It it self--evident that for narrow resonances the effects
related to changes of the width become enhanced. Thus the
observed mass dependence is understandable. It is clear also
that the initial state radiation, not included in our
calculations, dilutes sharp resonances and significantly
reduces effects due to changes of the width.
Therefore, in our opinion, for $m_t$
around $150$~GeV and larger theoretical uncertainties
introduced by model assumptions of the present paper, or those
introduced in \cite{Sumino}, may be neglected.
In particular
the position of $1S$ peak is rather insensitive to model
assumptions on the width. The normalization of the
cross section at the peak is sensitive to these assumptions
if the top mass is in the lower range of the allowed region.
Thus a rigorous (not model)
calculation including the effects discussed in this paper
may prove to be indispensable
if the top mass is around 120 GeV and  one aims at a
high precision study based on the absolute normalization of
the cross section at $1S$ peak.\\

However, we would like to stress once again that considering
only phase space suppression one overestimates effects of the
momentum dependent width. It is likely that rigorous calculations
will give results similar to those of Model II, which are quite
close to the predictions obtained using the constant width.

\section{Summary}

In this paper we have studied the influence of the
energy and momentum dependent
width $\Gamma(p,E)$
on $t-\bar t$ production near the energy threshold.
The position of the $1S$ peak in the total cross section as well
as the maximum of $\frac{{\rm d}\sigma}{{\rm d}p}$ are rather
insensitive to model assumptions related to $\Gamma(p,E)$.
The total cross section
$\sigma(e^+e^- \to t\bar t\,)$
around the $1S$ peak reflects the effective change of the width
introduced by $\Gamma(p,E)$.
The effects of the momentum dependent width are likely to be much
smaller than expected from the arguments based on the reduction
of the available phase space.
Further dilution of the $1S$ peak due to initial
state radiation results in additional reduction of the
differences between the results for the momentum dependent
and constant widths.
The energy spectra of $W$ bosons are slightly
narrower, because the momentum dependence in $\Gamma(p,E)$
results in suppression of the large $p$ components of $t-\bar t$
wave functions.
\newpage
\vskip 1.5cm
\par\noindent{\Large\bf Acknowledgements:}\\
\vskip 0.5cm
\noindent
We thank I.~Bigi, G.~Jikia, V.~Khoze, M.~Peskin and H.~Pilkuhn
for discussions. We are especially indebted to Johann
K\"uhn for encouragement, reading this manuscript,
and helpful discussions and remarks.\\
Work partly supported by BMFT, and by
Polish Committee for Scientific Research
(KBN) under Grants No.~20-38-09101 and 22-37-29102.


\sloppy
  \def\thebibliography#1{{{\noindent {\Large\bf References}}}
  \list
   {[\arabic{enumi}]}{\settowidth\labelwidth{[#1]}\leftmargin
   \labelwidth
   \advance\leftmargin\labelsep
   \usecounter{enumi}}
   \sloppy
   \sfcode`\.=1000\relax}
  \let\endthebibliography=\endlist
\def\app#1#2#3{{\it Act. Phys. Pol. }{\bf B #1} (#2) #3}
\def\jl#1#2#3{{\it JETP Lett. }{\bf #1} (#2) #3}
\def\lhc{Proc. LHC Workshop, CERN 90-10}
\def\npb#1#2#3{{\it Nucl. Phys. }{\bf B #1} (#2) #3}
\def\plb#1#2#3{{\it Phys. Lett. }{\bf B #1} (#2) #3}
\def\anph#1#2#3{{\it Ann. of Phys. }{\bf#1} (#2) #3}
\def\prd#1#2#3{{\it Phys. Rev. }{\bf D #1} (#2) #3}
\def\prl#1#2#3{{\it Phys. Rev. Lett. }{\bf #1} (#2) #3}
\def\prc#1#2#3{{\it Phys. Reports }{\bf C #1} (#2) #3}
\def\cpc#1#2#3{{\it Comp. Phys. Commun. }{\bf #1} (#2) #3}
\def\nim#1#2#3{{\it Nucl. Inst. Meth. }{\bf #1} (#2) #3}
\def\pr#1#2#3{{\it Phys. Reports }{\bf #1} (#2) #3}
\def\sovnp#1#2#3{{\it Sov. J. Nucl. Phys. }{\bf #1} (#2) #3}
\def\jet#1#2#3{{\it JETP Lett. }{\bf #1} (#2) #3}
\def\zpc#1#2#3{{\it Z. Phys. }{\bf C #1} (#2) #3}

\end{document}